# COVIDHunter: COVID-19 pandemic wave prediction and mitigation via seasonality aware modeling

Mohammed Alser    Jeremie S. Kim    Nour Almadhoun Alserr    Stefan W. Tell    Onur Mutlu

*ETH Zurich, Switzerland*

**ABSTRACT**

Early detection and isolation of COVID-19 patients are essential for successful implementation of mitigation strategies and eventually curbing the disease spread. With a limited number of daily COVID-19 tests performed in every country, simulating the COVID-19 spread along with the potential effect of each mitigation strategy currently remains one of the most effective ways in managing the healthcare system and guiding policy-makers. We introduce COVIDHunter, a flexible and accurate COVID-19 outbreak simulation model that evaluates the current mitigation measures that are applied to a region, predicts COVID-19 statistics (the daily number of cases, hospitalizations, and deaths), and provides suggestions on what strength the upcoming mitigation measure should be.

The key idea of COVIDHunter is to quantify the spread of COVID-19 in a geographical region by simulating the average number of new infections caused by an infected person considering the effect of external factors, such as environmental conditions (e.g., climate, temperature, humidity), different variants of concern, vaccination rate, and mitigation measures. Using Switzerland as a case study, COVIDHunter estimates that we are experiencing a deadly new wave that will peak on 26 January 2022, which is very similar in numbers to the wave we had in February 2020. The policy-makers have only one choice that is to increase the strength of the currently applied mitigation measures for 30 days. Unlike existing models, the COVIDHunter model accurately monitors and predicts the daily number of cases, hospitalizations, and deaths due to COVID-19.

Our model is flexible to configure and simple to modify for modeling different scenarios under different environmental conditions and mitigation measures. We release the source code of the COVIDHunter implementation at https://github.com/CMU-SAFARI/COVIDHunter and show how to flexibly configure our model for any scenario and easily extend it for different measures and conditions than we account for.

## 1. INTRODUCTION

*Coronavirus disease 2019* (COVID-19) is caused by SARS-CoV-2 virus, which has rapidly spread to nearly every corner of the globe and has been declared a pandemic in March 2020 by the World Health Organization (WHO) [1]. As of November 2021, only about 40% of the entire world population is fully vaccinated and their protection wanes after a few months [2]. Until an effective drug or vaccination is made widely available to everyone, early detection and isolation of COVID-19 patients remain essential for effectively curbing the disease spread [3]. Regardless of the availability and affordability of COVID-19 testing, it is still extremely challenging to detect and isolate COVID-19 infections at early stages [4], [5]. Simulating the spread of COVID-19 has the potential to mitigate such challenges, help to better manage the healthcare system, and provide guidance to policy-makers on the effectiveness of various (current, planned, or discussed) mitigation measures. To this end, many COVID-19 simulation models are proposed [6]–[10], some of which are announced to *assist* in decision-making for policy-makers in countries such as the United Kingdom (ICL [9]), United States (IHME [10]), and Switzerland (IBZ [11]).

These models tend to follow one of two key approaches. The first approach evaluates the current actual epidemiological situation by accounting for reporting delays and under-reporting (uncertainty) due to inefficiencies such as a low number of COVID-19 tests. This approach is taken by the IBZ [11], LSHTM [7], and [8] models and is *not* mainly used for prediction purposes as it reflects the epidemiological situation with about two weeks of time delay (due to its dependence on observed COVID-19 reports). The IBZ model [11] estimates the daily reproduction number, $R$, of SARS-CoV-2 from observed COVID-19 incidence time series data after accounting for reporting delays and under-reporting using the numbers of confirmed hospitalizations and deaths. The $R$ number describes how a pathogen spreads in a particular population by quantifying the average number of new infections caused by each infected person at a given point in time [12]. The LSHTM model [7] adjusts the daily number of observed COVID-19 cases by accounting for under-reporting (uncertainty) using both deaths-to-cases ratio estimates and correcting for delays between case confirmation (i.e., laboratory-confirmed infection) to death.

The second approach evaluates the current epidemiological situation and predicts the future epidemiological situation by simulating the COVID-19 outbreak and considering the effects of mitigation measures. This approach, taken by ICL [9] and IHME [10] models, usually suffers from two main drawbacks. The first drawback is that they require a large number of country-specific assumptions and input parameters (e.g., mobility rates, age- and country-specific data on demographics, patterns of social contact, and hospital availability) as it does not rely on the observed (laboratory-confirmed) number of cases for each region in simulation. For example, ICL [9] model requires input parameters such as the daily number of confirmed deaths, IFR, mobility rates from Google, age- and country-specific data on demographics, patterns of social contact, and hospital availability. This model makes three key assumptions: 1) age-specific IFRs observed in China and Europe are the same across every country, 2) the number of confirmed deaths is equal to the true number of COVID-19 deaths, and 3) the change in transmission rates is a function of average mobility trends. Another example is the IHME [10] model, which requires input parameters such as testing rates, mobility, social distancing policies, population density, altitude, smoking rates, self-reported contacts, and mask use. This model makes two key assumptions: 1) the infection fatality rate (IFR), which indicates the rate of people that die from the infection is taken using data from the Diamond Princess Cruise ship and New Zealand and 2) the decreasing fatality rate is reflective of increased testing rates (identifying higher rates of asymptomatic cases). The second drawback is the lack of awareness about environmental conditions of the subject region, they usually provide inaccurate estimates especially during the winter months [13]. Several related viral infections, such as the Influenza virus, human coronavirus, and human respiratory, already show notable seasonality (showing peak incidences during only the winter (or summer) months) [14], [15]. There are currently several studies that demonstrate the strong dependence of the transmission of SARS-CoV-2 virus on one or more environmental conditions, even after controlling (isolating) the impact of mitigation measures and behavioral changes that reduce contacts [16]–[21].

To our knowledge, there is currently no model capable of accurately monitoring the current epidemiological situation and predicting future scenarios while considering a reasonably low number of parameters and accounting for the effects of environmental conditions (**Table 1**).

**Table 1. Comparison to other models used to inform government policymakers, as of January 2021**

| Model | Open Source | Well Documented[#] | Accounting for Seasonality | Low Number of Parameters | Reported COVID-19 Statistics |
|---|---|---|---|---|---|
| COVIDHunter (this work) | ✓ | ✓ | ✓ | ✓ | ✓ ($R$, cases, hospitalizations, and deaths) |
| IBZ [11] | ✓ | ✗ | ✗ | ✓ | ✗ (only $R$) |
| LSHTM [7] | ✓ | ✗ | ✗ | ✓ | ✗ (only cases) |
| ICL [9] | ✓ | ✓ | ✗ | ✗ | ✓ ($R$, cases, hospitalizations, and deaths) |
| IHME [10] | ✓* | ✗ | ✗ | ✗ | ✗ (cases, hospitalizations, and deaths) |

\* The available packages are configured only for the IHME infrastructure.
\# Based on the documentation available on each model's GitHub page (all models are available on GitHub).

Our **goal** in this work is to develop and validate such a COVID-19 outbreak simulation model. To this end, we introduce *COVIDHunter*, a simulation model that evaluates the current mitigation measures (i.e., non-pharmaceutical intervention or NPI) that are applied to a region and provides insight into what strength the upcoming mitigation measure should be and for how long it should be applied, while considering the potential effect of environmental conditions. Our model accurately forecasts the numbers of infected and hospitalized patients, and deaths for a given day, as validated on historical COVID-19 data (after accounting for under-reporting). The **key idea** of COVIDHunter is to quantify the spread of COVID-19 in a geographical region by calculating the daily reproduction number, $R$, of COVID-19 and scaling the reproduction number based on changes in mitigation measures, environmental conditions, different variants of concern, and vaccination rate. The $R$ number changes during the course of the pandemic due to the change in the ability of a pathogen to establish an infection during a season and mitigation measures that lead to a lower number of susceptible individuals. COVIDHunter simulates the entire population of a region and assigns *each* individual in the population to a stage of the COVID-19 infection (e.g., from being healthy to being short-term immune to COVID-19) based on the scaled $R$ number. COVIDHunter requires *only* three input parameters, two of which are time-varying parameters, to calculate the $R$ number, which provides four key advantages: 1) allowing flexible (easy-to-adjust) configuration of the model input parameters for different scenarios and different geographical regions, 2) enabling short simulation execution time and simpler modeling, 3) enabling easy validation/correction of the model prediction outcomes by adjusting fewer variables, and 4) being extremely useful and powerful especially during the early stages of a pandemic as many of the parameters are unknown. Whenever applicable, we compare the simulation output of our model to that of four state-of-the-art models currently used to inform policy-makers, IBZ [11], LSHTM [7], ICL [9], and IHME [10].

## 2. MATERIALS AND METHODS

The COVIDHunter model employs a four-stage approach to simulate the COVID-19 outbreak (**Figure 1**). 1) Predicting the daily reproduction number, the average number of new daily infections caused by each infected person. 2) COVIDHunter simulates the entire population of a region and labels each individual according to different stages of the COVID-19 infection timeline. Each stage has a different degree of infectiousness and contagiousness. The model simulates these stages for each individual to maintain accurate predictions. 3) Predicting the number of daily cases based on our population simulation. 4) Predicting the number of daily deaths and hospitalizations based on both the predicted number of cases and the $R$ number. All input parameters to our model are fully configured based on either existing research findings or user-defined values.

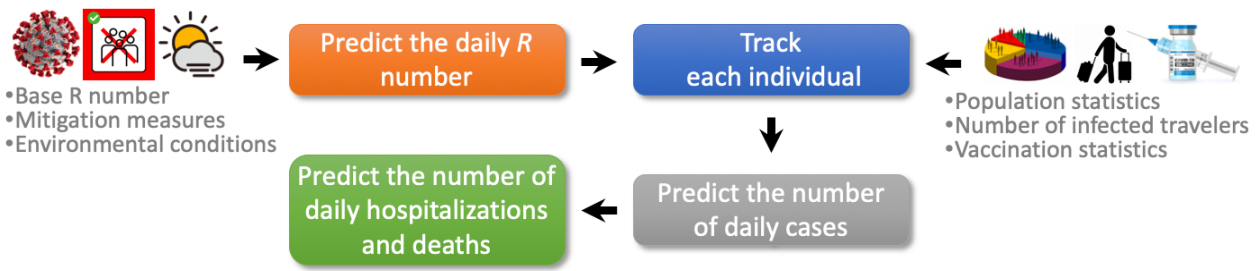

Figure 1. Proposed COVIDHunter model for simulating COVID-19 outbreak.

### 2.1 Predicting the reproduction number

One of the most challenging factors in predicting the spread of COVID-19 is to quantify the daily reproduction number ($R$) due to the large number of factors affecting its value and various viral genetic variations. The $R$ number is *directly* affected by a large number of factors [12], such as 1) the transmissibility of the virus variant of concern, 2) the strength of the mitigation measures, 3) weather factors (e.g., temperature), 4) air pollutants, 5) population density, and many more. The coronavirus genome can also exhibit rapid genetic changes in its nucleotide sequence [22], [23]. This genetic diversity affects the virus virulence, infectivity, transmissibility, and evasion of the host immune responses [23], [24]. To provide accurate predictions of the reproduction number, the COVIDHunter model considers *only* three key factors for predicting the $R$ number: 1) different transmissibility rates of infection into a susceptible host population for each SARS-CoV-2 variant, 2) mitigation measures (e.g., lockdown, social distancing, and isolating infected people), and 3) environmental conditions (e.g., air temperature). We choose these three main factors for two reasons: 1) they have a large impact on the $R$ number [3], [14], [15], 2) the mitigation measure and the environmental conditions can represent almost any other factor that affects the $R$ number (e.g., high population density can be thought of as a weaker mitigation measure). The COVIDHunter model allows for *directly* leveraging existing models that study the effect of *only* mitigation measures (or *only* environmental conditions) on the spread of COVID-19. Our model calculates the time-varying $R$ number using **Equation 1** as follows:

$$R(t) = R0 * (1 - M(t)) * C_e(t) \quad (1)$$

where $R0$ is the base reproduction number for the virus variant of concern, $M(t)$ is the mitigation coefficient for the given day $t$, and $C_e(t)$ is the environmental coefficient for the given day $t$. The $R0$ number quantifies the transmissibility of infection into a susceptible host population by calculating the expected average number of new infections caused by an infected person in a population with no prior immunity to a specific virus or variant (as a pandemic virus is by definition novel to all populations). Hence, the $R0$ number represents the transmissibility of an infection at only the beginning of the outbreak assuming the population is not protected via vaccination. Unlike the $R$ number, the $R0$ number is a fixed value and it does not depend on time. The $R0$ number for each SARS-CoV-2 variant can be obtained from several existing studies (such as in [25]–[28]) that estimate it by modeling contact patterns during the first wave of the pandemic.

The mitigation coefficient ($M(t)$) applied to the population is a time-dependent variable and it has a value between 0 and 1, where 1 represents the strongest mitigation measure and 0 represents no mitigation measure applied. In different countries, mitigation measures take different forms, such as social distancing, self-isolation, school closure, banning public events, and complete lockdown. These measures exhibit significant heterogeneity and differ in timing and intensity across countries [9]. *The Oxford Stringency Index* [29] maintains a twice-weekly-updated index that represents the severity of nine mitigation measures that are applied by more than 160 countries. Another study [30] estimates the effect of *only* seven mitigation measures on the $R$ number in 41 countries. We can *directly* leverage such studies for calculating the mitigation coefficient on a given day.

The environmental coefficient ($C_e(t)$) is a time-dependent variable representing the effect of external environmental factors on the spread of COVID-19 and it has a value between 0 and 2. Several studies have demonstrated increased infectiousness by a country-dependent fixed-rate with each 1 °C fall in daytime temperature [16], [17]. Another study supports the same temperature-infectiousness relationship, but it also finds that before applying any mitigation measures, a one-degree drop in relative humidity shows increased infectiousness by a rate lower (2.94× less) than that of temperature [19]. Another study follows a simple way of modeling the effect of seasonality on COVID-19 transmission using a sinusoidal function with an annual period [20]. One of the most comprehensive studies that spans more than 3700 locations around the world is *HARVARD CRW* (or CRW in short) [21]. It finds the statistical correlation between the relative changes in the $R$ number and both weather conditions and air pollution after controlling the impact of mitigation measures. Our model enables applying *any of these studies* as we experimentally demonstrate in **Section 3**. In our experiments, we choose two main approaches for setting the value of the time-varying environmental coefficient variable ($C_e(t)$). 1) The first approach is to perform statistical analysis for the relationship between the daily number of COVID-19 cases and average daytime temperature in Switzerland. 2) The second approach is to apply the *HARVARD CRW* [21] (referred to as *CRW*). Next, we explain the first approach in detail.

**2.2 Statistical relationship between temperature and number of COVID-19 cases**

To calculate the environmental coefficient, we explore the relationship between the daily new confirmed COVID-19 case counts or death counts and temperature in Switzerland. We obtain the daily number of confirmed COVID-19 cases and deaths in Switzerland from official reports of the Federal Office of Public Health (FOPH) in Switzerland [31] starting from March 2020 until January 2021. We obtain the air temperature data from the Federal Office of Meteorology and Climatology (MeteoSwiss) in Switzerland [32]. We calculate the daily average air temperature during the same time period (March 2020 to December 2020) for all the 26 cantons in Switzerland. To evaluate the correlation between the temperature data and the number of daily confirmed COVID-19 cases or the daily counts of death, we use a generalized additive model (GAM). GAM is usually used to calculate the linear and non-linear regression models between meteorological factors (e.g., temperature, humidity) with COVID-19 infection and transmission [16], [17], [33].

Our analyses are performed with R software version 4.0.3, where *p*−value < 0.05 is considered statistically significant. Our model attempts to represent the linear behavior of the growth curve of the counts of the new confirmed cases or deaths in Switzerland. Therefore, we can test the hypothesis of whether there is a significant negative correlation between the COVID-19 confirmed daily case or death counts and temperature. The results demonstrate a significant negative correlation between temperature and COVID-19 daily case and death counts. Specifically, the relationship is linear for the average temperature in the range from 1-26°C. Based on **Figure 2**, we make two key observations. 1) For each 1°C rise in temperature, there is a 3.67% (*t*-value = 3.244 and *p*-value = 0.0013) decrease in the daily number of COVID-19 confirmed cases (**Figure 2(a)**). 2) For each 1°C rise in temperature, there is a 23.8% decrease in the daily number of COVID-19 deaths (*t*-value = 9.312 and *p*-value = 0.0), as shown in **Figure 2(b)**. The statistical analysis can be reproduced using the following script https://github.com/CMU-SAFARI/COVIDHunter/tree/main/TemperatureSensitivityStudy.

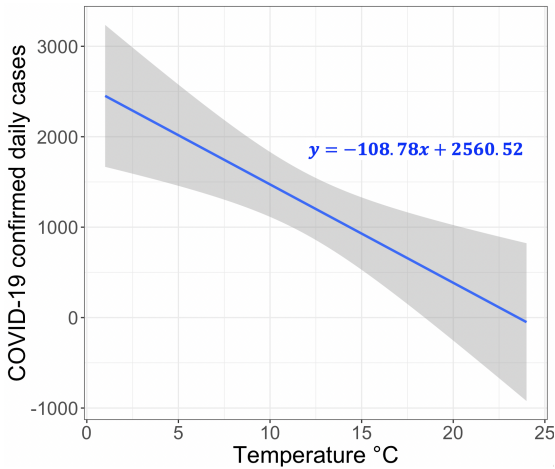
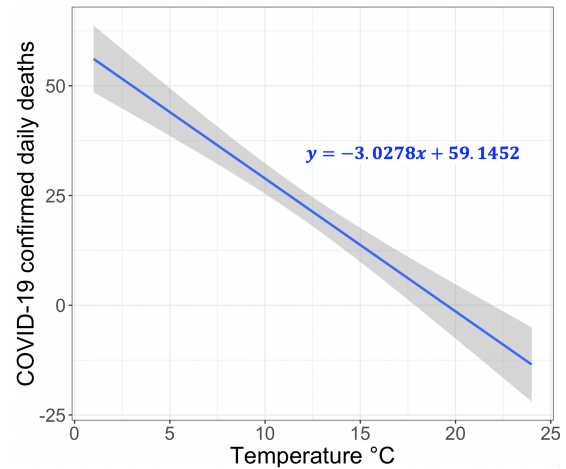

(a)                        (b)

**Figure 2.** Correlation between temperature and COVID-19 confirmed (a) case count and (b) death count in 26 cantons of Switzerland.

### 2.3 Labeling each individual in the subject population according to different stages of the COVID-19 infection timeline.

COVIDHunter tracks the number of infected and uninfected persons over time by clustering the population into eight main categories: HEALTHY, VACCINATED, INFECTED, CONTAGIOUS, HOSPITALIZED, IMMUNE, DEAD, and INFECTED TRAVELERS (**Figure 3**). The model initially considers the entire population as uninfected (i.e., HEALTHY). For each simulated day, the COVIDHunter model decides which persons will have immunity to infection due to vaccination (i.e., VACCINATED) based on input data. For the unvaccinated persons, the model calculates the $R$ value using **Equation 1** (**Section 2.1**) and decides how many persons can be infected (i.e., INFECTED) during each simulated day. Our modeling approach considers multiple virus strains/variants by calculating multiple $R$ numbers, each of which represents a different virus strain/variant. The day when the first case of infection (caused by a variant of concern) in a population introduced is defined by the user. For each newly infected person (INFECTED), the model maintains a counter that counts the number of days from being infected to being contagious (CONTAGIOUS). Several COVID-19 case studies show that *presymptomatic* transmission can occur 1–3 days before symptom onset [34], [35]. COVID-19 patients can develop symptoms mostly after an incubation period of 1 to 14 days (the median incubation period is estimated to be 4.5 to 5.8 days) [4], [5]. We calculate the number of days of being contagious after being infected as a random number with a Gaussian distribution that has user-defined lowest and highest values. Each contagious person may infect $N$ other persons depending on mobility, population density, number of households, and several other factors [36]. We calculate the value of $N$ to be a random number with a Gaussian distribution that has the lowest value of 0 and the highest value determined by the user. If $N$ is greater than the $R$ number (i.e., the target number of infections for that day has been reached), further infections are curtailed preventing overestimation of $N$ by infecting *only* $R$ persons.

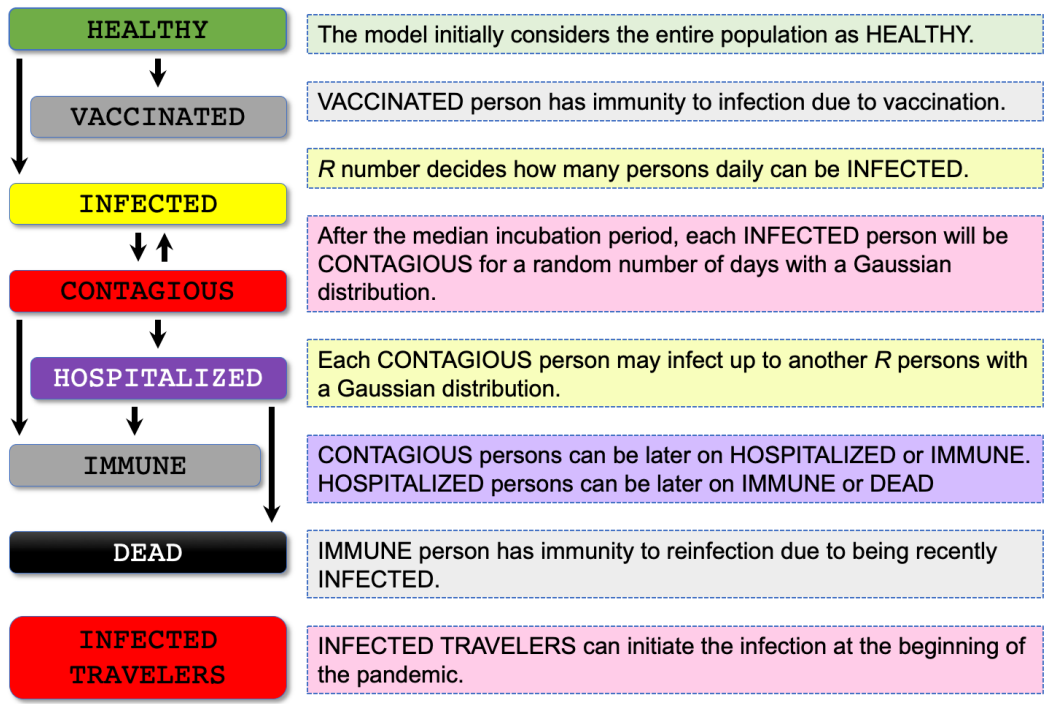

**Figure 3. Proposed population clustering algorithm** for assigning each individual in the population of concern to a stage of the COVID-19 infection timeline. The COVIDHunter model makes eight main clusters: HEALTHY, VACCINATED, INFECTED, CONTAGIOUS, HOSPITALIZED, IMMUNE, DEAD, and INFECTED TRAVELERS.

Once the contagious person infects the desired number of susceptible persons, the status of the contagious person becomes IMMUNE or HOSPITALIZED. The IMMUNE status indicates that the person has immunity to reinfection due to either vaccination or being recently infected [37], [38]. The HOSPITALIZED person can be later on IMMUNE or DEAD. There are currently two key approaches for calculating the estimated number of both hospitalizations and deaths due to COVID-19: 1) using historical statistical probabilities, each of which is unique to each age group in a population [39], [40] and 2) using historical COVID-19 hospitalizations-to-cases and deaths-to-cases ratios [41]. We choose to follow the second approach as it does *not* require 1) clustering the population into age-groups and 2) calculating the risk of each individual using the given probability, which both affect the complexity of the model and the simulation time. As the *true* number of cases is unknown due to both lack of population-scale testing and asymptomatic cases [42], [43], it is extremely difficult to make accurate estimates of the *true* number of COVID-19 hospitalizations and deaths. As such, we assume a fixed multiplicative relationship between the number of laboratory-confirmed cases and the *true* number of cases. We use user-defined correction coefficients (we refer to them as *certainty rate levels*) to account for such a multiplicative relationship. A certainty rate of, for example, 50% means that the *true* number of COVID-19 cases is actually *double* that calculated by COVIDHunter.

Our model also simulates the effect of infected travelers (i.e., INFECTED TRAVELERS) on the value of $R$. These travelers (e.g., daily cross-border commuters within the European Union) can initiate the infection(s) at the beginning of the pandemic. If such infected travelers are absent (due to, for example, emergency lockdown) from the target population, the virus would die out once the value of $R$ decreases below 1 for a sufficient period of time. The percentage of incoming infected travelers is *not* affected by the changes in the local mitigation measures nor the environmental conditions, as these travelers were already infected abroad.

**2.4 Predicting the Number of COVID-19 Cases**

The COVIDHunter model assigns each individual in the entire population of a region to a stage of the COVID-19 infection timeline. Using this assignment, our model predicts the *daily* number of COVID-19 cases for a given day $t$, as follows:

$$Daily\_Cases(t) = \sum_{n=0}^{T_{INF}(t)} N(n) + \sum_{m=0}^{U_{CON}(t)} N(m) \qquad (2)$$

where $T_{INF}$ is the daily number of infected travelers that is a user-defined variable, $N(\ )$ is a function that calculates the number of persons to be infected by a given person as a random number with a Gaussian distribution, and $U_{CON}$ is the daily number of contagious persons calculated by our model.

**2.5 Predicting the number of COVID-19 hospitalizations and deaths**

The number of COVID-19 hospitalizations for a given day, $t$, can be calculated as follows:

$$Daily\_Hospitalizations(t) = Daily\_Cases(t) * X * C_X \qquad (3)$$

where $Daily\_Cases(t)$ is calculated using **Equation 2** and $X$ is the hospitalizations-to-cases ratio that is calculated as the average of daily ratios of the number of COVID-19 hospitalizations to the laboratory-confirmed number of COVID-19 cases. As the *true* number of cases is unknown due to both lack of population-scale testing and asymptomatic cases [42], [43], it is extremely difficult to make accurate estimates of the *true* number of COVID-19 hospitalizations. As such, we assume a fixed multiplicative relationship between the number of laboratory-confirmed cases and the *true* number of cases. We use the user-defined correction coefficient, $C_X$, of the hospitalizations-to-cases ratio to account for such a multiplicative relationship. The number of COVID-19 deaths for a given day $t$ can be calculated as follows:

$$Daily\_Deaths(t) = Daily\_Cases(t) * Y * C_Y \qquad (4)$$

where $Daily\_Cases(t)$ is calculated using **Equation 2** and $Y$ is the deaths-to-cases ratio, which is calculated as the average of daily ratios of the number of COVID-19 deaths to the number of COVID-19 laboratory-confirmed cases. The observed number of COVID-19 deaths can still be less than the *true* number of COVID-19 deaths due to, for example, under-reporting. We use the user-defined correction coefficient, $C_Y$, to account for the under-reporting. One way to find the *true* number of COVID-19 deaths is to calculate the number of excess deaths. The number of excess deaths is the difference between the observed number of deaths during a time period and the expected (based on historical data) number of deaths during the same time period. For this reason, $C_Y$ may not necessarily be equal to $C_X$.

## 3. RESULTS

We evaluate the daily 1) $R$ number, 2) mitigation measures, and 3) numbers of COVID-19 cases, hospitalizations, and deaths. We compare the predicted values to their corresponding observed values and that of four state-of-the-art models, ICL [9], IHME [10], IBZ [11], and LSHTM [7], whenever possible. We provide a comprehensive treatment of *all* datasets, models, and evaluation results with different model configurations in the Supplementary Materials and on GitHub page of COVIDHunter, https://github.com/CMU-SAFARI/COVIDHunter. We also provide all parameter values used for running COVIDHunter and different scripts for reproducing the experimental evaluation performed in this work on our GitHub page, https://github.com/CMU-SAFARI/COVIDHunter/tree/main/Reproduce-Switzerland-Case-Study-Results . We provide below our prediction run for the period of 20 November 2021 until February 2022, which was carried out on 20 November 2021. We provide another prediction run for the period from 19 April 2021 until 1 June 2021 in the supplementary materials, which were carried out on 19 April 2021. We provide a prediction run in [44] for the period from 22 January 2021 until 22 February 2021, which was carried out on 22 January 2021. We also provide a comprehensive analysis of the COVID-19 statistics provided by ICL [9], IHME [10], IBZ [11], and LSHTM [7] from the beginning of the COVID-19 outbreak (February 2020) until April 2021.

### 3.1 Determining the value of each variable in the equations

We use Switzerland as a use-case for all the experiments. However, our model is not limited to any specific region as the parameters it uses are completely configurable. To predict the $R$ number, we use **Equation 1** that requires three key variables. We set the base reproduction numbers, $R0$, for two main variants, the Delta variant and its ancestral strain, of SARS-CoV-2 in Switzerland as 5 and 2.7, respectively, as shown in [25], [26], [45]. The recent Omicron variant was not circulating during the study. We set the first day for the Delta variant to be injected into the population as 19 June 2021 based on the governmental data [46]. We set the first day of vaccination availability in Switzerland as 28 February 2021, with a vaccination rate of 0.28 per day based on governmental data [46]. We change the daily mitigation coefficient, $M(t)$, value based on the ratio of number of confirmed hospitalizations to the number of confirmed cases with two certainty rate levels of 100% and 50%, as we explain in detail in **Section 3.2**. This helps us to take into account uncertainty in the observed number of COVID-19 cases, hospitalizations, and deaths. We set the minimum and maximum incubation time for SARS-CoV-2 as 1 and 5 days, respectively, as 5-day period represents the median incubation period worldwide [4], [5]. We set the population of Switzerland to 8654622. We empirically choose the values of $N$, the number of travelers, and the ratio of the number of infected travelers to the total number of travelers to be 25, 100, and 15%, respectively.

### 3.2 Model validation

We can validate our model using two key approaches. 1) Comparing the daily $R$ number predicted by our model (using **Equation 1**) with the daily reported official $R$ number for the same region. 2) Comparing the daily number of COVID-19 cases predicted by our model (using **Equation 2**) with the daily number of laboratory-confirmed COVID-19 cases. We decide to use a combination of reported numbers of cases, hospitalizations, and deaths to validate our model for three main reasons. 1) The $R$ number is calculated as, for example, the ratio of the number of cases for a week (7-day rolling average) to the number of cases for the preceding week. Adjusting the parameters of our model to fit the curve of the number of confirmed cases is likely to be highly uncertain. 2) The reported daily reproduction number by authorities of Switzerland usually excludes the values for the last 14 days, which makes the validation based on the reproduction number more challenging. 3) As of 2022, we have already witnessed more than two years of the pandemic, which provide us with several observations and lessons. The most obvious source of uncertainty, affecting *all* models, is that the *true* number of persons that are previously infected or currently infected is *unknown* [47]. However, the publicly-available number of COVID-19 hospitalizations and deaths can provide more reliable data.

We validate our model using three key steps. 1) We leverage the more reliable data of reported number of hospitalizations (or deaths) to estimate the *true* number of COVID-19 cases using the ratio of number of laboratory-confirmed hospitalizations (or deaths) to the number of laboratory-confirmed cases during the second wave of the COVID-19 pandemic. We assume that the COVID-19 statistics during the second wave is more accurate than that during the first wave because generally more testing is performed in the second wave. 2) We consider a multiplicative relationship between the *true* number of COVID-19 cases and that estimated in step 1. In our experimental evaluation, we use the *true* number of COVID-19 cases calculated using different multiplicative factor values (we refer to them as *certainty rate levels*) as a ground-truth for validating our model. A certainty rate of, for example, 50% means that the *true* number of COVID-19 cases is actually *double* that calculated in step 1. 3) We use our model to calculate both the daily $R$ number (**Equation 1**) and the number of COVID-19 cases (**Equation 2**). We fix the two terms of **Equation 1**, $R0$ and $C_e$, using publicly-available data for a given region and change the third term, $M$, until we fit the curve of the number of cases predicted by our model to the ground-truth plot calculated in step 2.

**3.3 Evaluating the expected number of COVID-19 cases for model validation**

As the exact true number of COVID-19 cases remains unknown (due to, for example, lack of population-scale COVID-19 testing), we expect the true number of COVID-19 cases in Switzerland to be higher than the observed (laboratory-confirmed) number of cases. We calculate the expected true number of cases based on both numbers of deaths and hospitalizations, as we explain in **Section 3.2**. To account for the possible missing number of COVID-19 deaths, we consider the excess deaths instead of observed deaths. We calculate the excess deaths as the difference between the 5-year average of weekly deaths and the observed weekly number of deaths in both 2020 and 2021. We find that $X$ (hospitalizations-to-cases ratio) and $Y$ (deaths-to-cases ratio, using excess death data) to be 3.75% and 2.441%, respectively, during the second wave of the pandemic in Switzerland. We choose the second wave to calculate the values of $X$ and $Y$ as Switzerland has increased the daily number of COVID-19 testing by $5.31 \times$ (21641/4074) on average compared to the first wave. We calculate the expected number of cases on a given day $t$ with certainty rate levels of 100% and 50% based on hospitalizations by dividing the number of hospitalizations at $t$ by $X$ and $X/2$, respectively, as we show in **Figure 4**. We apply the same approach to calculate the expected number of cases on a given day $t$ with certainty rate levels of 100% and 50% based on deaths using $Y$ and $Y/2$, respectively.

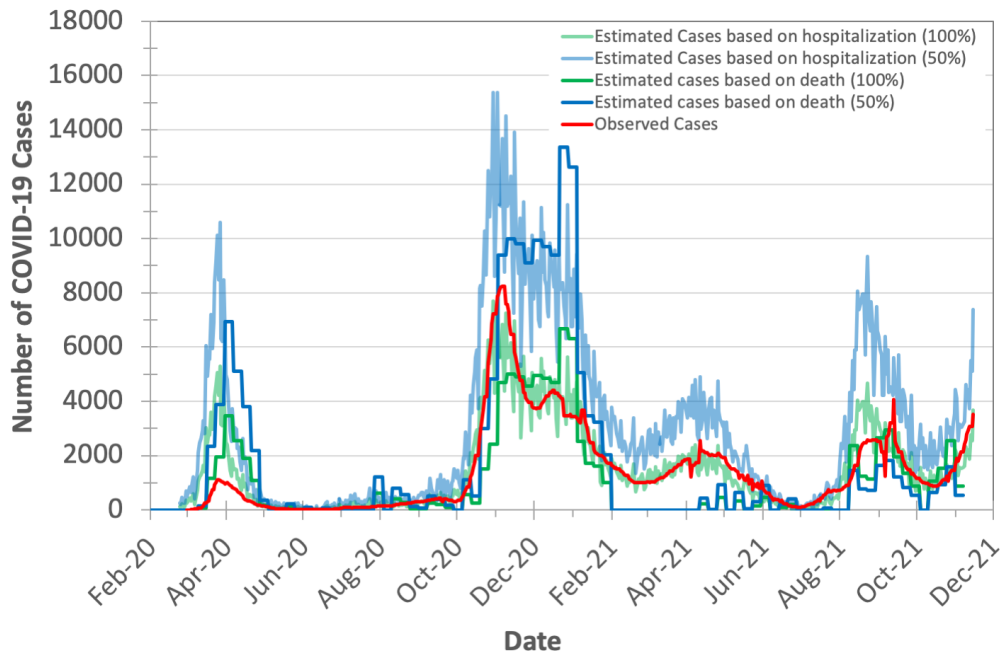

**Figure 4. Observed (officially reported) and expected number of COVID-19 cases in Switzerland during the years 2020 and 2021.** We calculate the expected number of cases based on both the hospitalizations-to-cases and deaths-to-cases ratios for the second wave. We assume two certainty rate levels of 50% and 100%.

Based on **Figure 4**, we make three key observations. 1) The plot for the expected number of cases calculated based on the number of deaths is shifted forward by 10-20 days (15 days on average) from that for the expected number of cases calculated based on the number of hospitalizations. This is due to the fact that each hospitalized patient usually spends some number of days in the hospital before dying of COVID-19. We do not observe a significant time shift between the plot of the expected number of cases calculated based on the number of hospitalizations and the plot of observed (laboratory-confirmed) cases. 2) The expected number of cases calculated based on the number of excess deaths is not reliable when the mass COVID-19 vaccination is kicked-off (after February 2021) as the number of deaths is quickly declined. 3) The expected number of cases calculated based on the number of hospitalizations is on average $2.7 \times$ higher than the expected number of cases calculated based on the number of excess deaths (after accounting for the 15-day shift) for the same certainty rate. This is expected as not all hospitalized patients die.

We conclude that the number of COVID-19 hospitalizations can be used reliably for estimating the true number of COVID-19 cases.

### 3.4 Evaluating the Predicted Number of COVID-19 Cases

We evaluate COVIDHunter's *predicted* daily number of COVID-19 cases in Switzerland. We compare the predicted numbers by our model to the observed numbers and those provided by two state-of-the-art models (ICL and IHME), as shown in **Figures 5(a),(b)**. We calculate the observed number of cases as the expected number of cases with a certainty rate level of 100% (as we discuss in **Section 3.3**). We use three default configurations for the prediction of the ICL model: 1) strengthening mitigation measures by 50%, 2) maintaining the same mitigation measures, and 3) relaxing mitigation measures by 50% which we refer to as ICL+50%, ICL, and ICL-50%, respectively, in **Figure 5**. We use the mean numbers reported by the IHME model. As we provide in **Section 2.2**, our statistical analysis shows that each 1°C rise in daytime temperature is associated with a 3.67% ($t$-value = -3.244 and $p$-value = 0.0013) decrease in the daily number of confirmed COVID-19 cases. We refer to this approach as Cases-Temperature Coefficient (CTC).

Based on **Figures 5(a),(b)**, we make three key observations. 1) Our model predicts that the peak (the highest number of COVID-19 cases) of the upcoming wave will be on 26 January 2022 (reaching up to 10,000 daily cases) and 31 December 2021 (reaching up to 44,800 daily cases and peaking up to 17 January 2022) for a certainty rate levels of 100% (**Figure 5(a)**) and 50% (**Figure 5(b)**), respectively, while maintaining the same strength of the current (20 November 2021) mitigation measures for 30 days. Both IHME and ICL models consider that the current number of COVID-19 cases in Switzerland shows a certainty rate level of 50% and the highest number of daily cases will be 10,000, but IHME and ICL models predict the peak of the upcoming wave to be on 26 January 2022 and 16 December 2021, respectively. 2) The number of COVID-19 cases reduces from 10,000 to 200 daily cases and from 44800 to 2,400 daily cases for a certainty rate levels of 100% (**Figure 5(a)**) and 50% (**Figure 5(b)**), respectively, within January 2021 if the mitigation measures that are applied nationwide in Switzerland are tightened by 50% (*M(t)* increases from 0.4 to 0.6 and from 0.3 to 0.5, respectively) for at least 30 days starting from 20 November to 20 December 2021. 3) Relaxing the mitigation measures before at least February 2022 can lead to a significant rise in the number of daily COVID-19 cases, reaching up to 43,500 as predicted by ICL and COVIDHunter (certainty rate levels of 100%) and up to 82,900 daily cases as predicted by COVIDHunter (certainty rate levels of 50%).

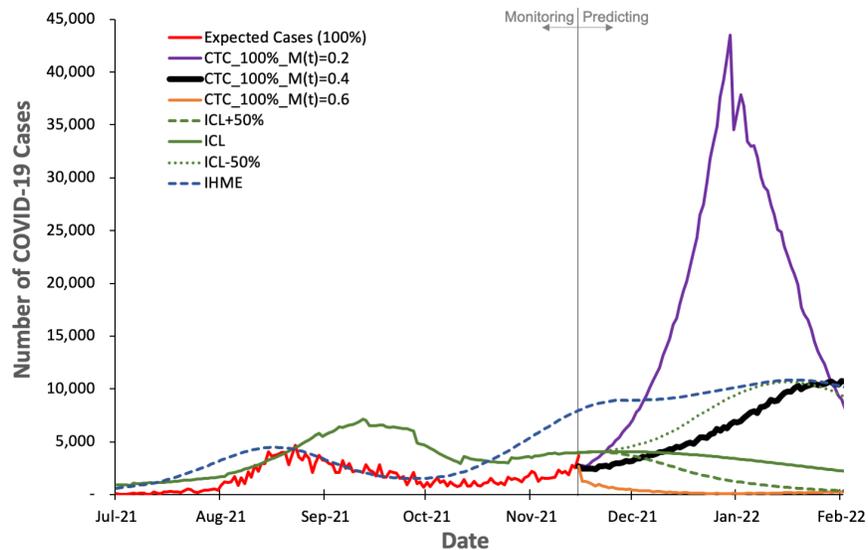

(a)

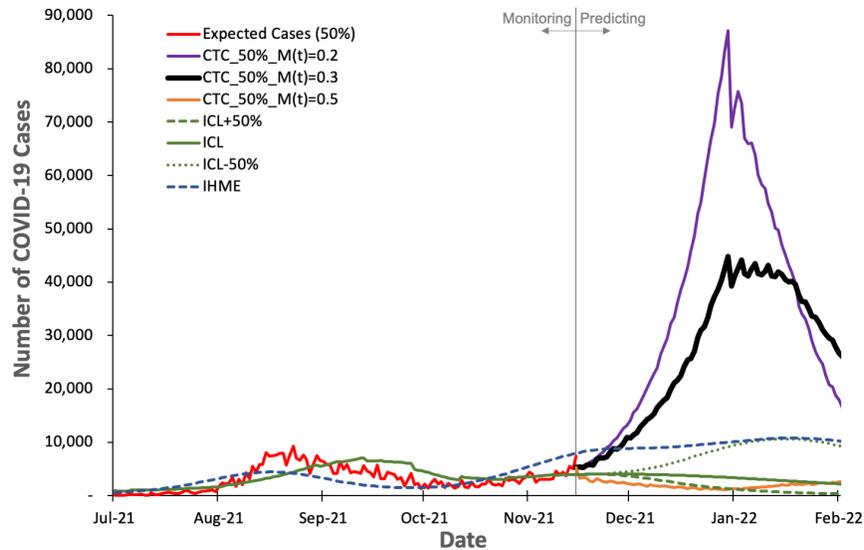

(b)

**Figure 5. Observed and predicted number of COVID-19 cases by our model and other two state-of-the-art models, ICL and IHME.** For COVIDHunter, we use CTC environmental condition approaches with two certainty rate levels of (a) 100% and (b) 50%. We show the prediction of COVIDHunter using three mitigation coefficient, $M(t)$, values, each of which is applied from 20 November to 20 December 2021. The predicted plot in a bold black line represents the situation when the mitigation measures applied before the prediction period remain the same.

### 3.5 Evaluating the Predicted Number of COVID-19 Hospitalizations and Deaths

We evaluate COVIDHunter's *predicted* daily number of COVID-19 hospitalizations and deaths in **Figure 6(a)** and **Figure 6(b)**. We use the observed official number of hospitalizations as is. We calculate the observed number of deaths as the number of excess deaths to account for uncertainty in reporting COVID-19 deaths. Using the number of cases calculated with Equation 2 and the observed number of hospitalizations and excess deaths (after accounting for 15-day shift, as we discuss in **Section 3.3** and **Figure 4**) during 1 August 2021 to 15 November 2021, we find $X$ (hospitalizations-to-cases ratio) and $Y$ (deaths-to-cases ratio, using excess death data) to be 1.508% and 0.498%, respectively. We choose the period from 1 August 2021 to 15 November 2021 for calculating the $X$ and $Y$ ratios to provide accurate predictions since the vaccination rate in Switzerland exceeds 50%, most of the risk groups received their second vaccination dose, and the Delta variant dominates the causes for COVID-19 cases [48].

Based on **Figures 6(a),(b)** we make four key observations. 1) IHME and ICL models consider that the current numbers of COVID-19 hospitalizations and deaths in Switzerland show a certainty rate level of 100%. 2) COVIDHunter and IHME show that the highest number of hospitalizations and deaths will be on 26 January 2022 (reaching up to 160 and 44 daily hospitalizations and deaths, respectively), which is a month and two weeks after that predicted by ICL for the number of hospitalizations and deaths, respectively. These predictions show that we will face a similar situation to the first wave we had in February 2020 if we maintain the same current mitigation measures. 3)

COVIDHunter and ICL models show that relaxing the current mitigation measures by 50% for a month (20 November to 20 December 2021) can increase the numbers of hospitalizations and deaths by up to 5.5x. They also show that tightening the current mitigation measures by 50% for a month (20 November to 20 December 2021) can reduce the numbers of hospitalizations and deaths by up to 3.9x.

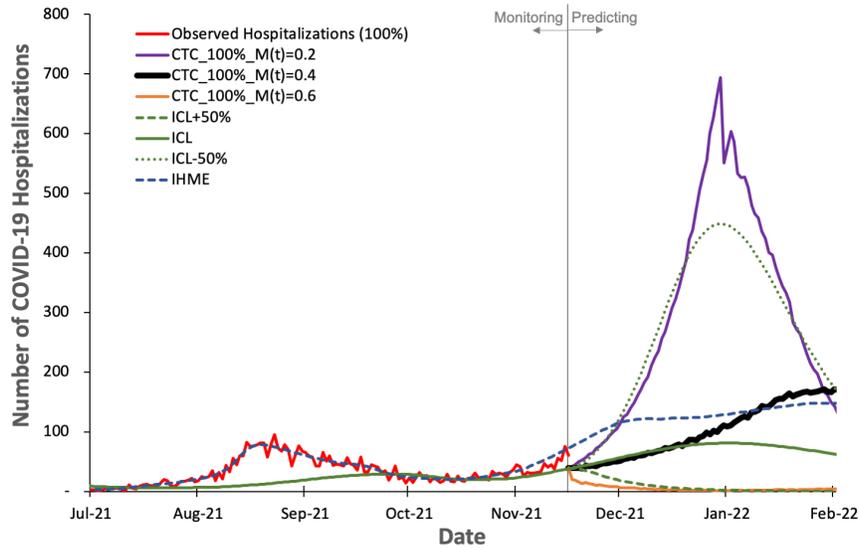

(a)

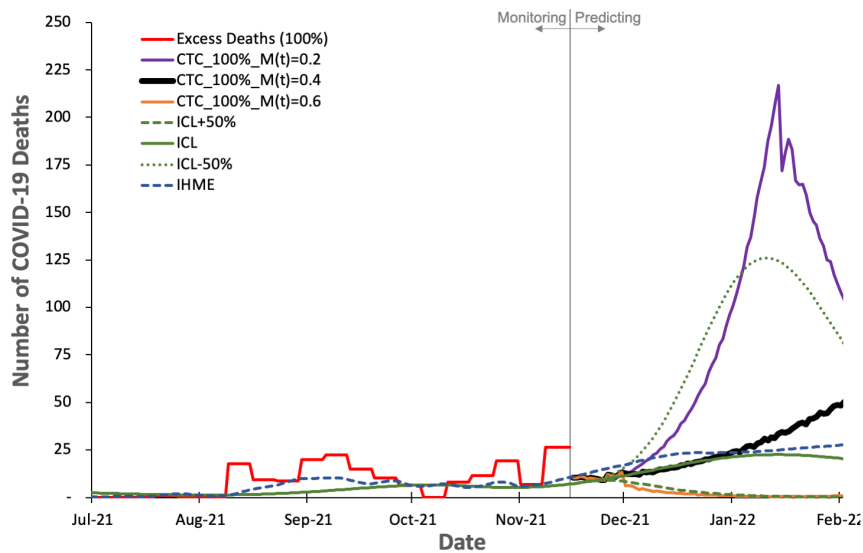

(b)

**Figure 6. Observed and predicted number of COVID-19 hospitalization and deaths by our model and other two state-of-the-art models, ICL and IHME.** For COVIDHunter, we use a certainty rate level of 100% for the numbers of (a) hospitalizations and (b) deaths, as IHME and ICL models tend to follow such a certainty rate. We show the prediction of COVIDHunter using three

mitigation coefficient, $M(t)$, values, each of which is applied from 20 November to 20 December 2021. The predicted plot in a bold black line represents the situation when the mitigation measures applied before the prediction period remain the same.

**3.6 Evaluating the Prediction Accuracy**

We evaluate the prediction accuracy of COVIDHunter, ICL, and IHME models using the real COVID-19 statistics that are published by the Federal Office of Public Health (FOPH) of Switzerland three months after performing the prediction. We evaluate the prediction accuracy for the number of cases, hospitalizations, and deaths due to COVID-19 in **Figures 7 and 8**.

     **Figure 7** shows the number of COVID-19 cases predicted by the three models and the real number (called "FOPH Cases" in **Figure 7**) of COVID-19 cases released by FOPH. FOPH usually does not report COVID-19 statistics during the weekends and thus we also show the 7-day rolling average numbers (called "Smoothed FOPH Cases" in **Figure 7**) of COVID-19 cases as provided by https://ourworldindata.org (referred to as "Smoothed data" in the supplementary materials, **Section S1**). We make four key observations. 1) COVIDHunter is the only model that is able to accurately predict the number of COVID-19 cases. Although COVIDHuner predicts the mitigation measures applied during November 2021 to be of strength 0.3 using a certainty rate level of 50% (**Figure 5(b)**), the mitigation measures have been already tightened during November and December 2021, as shown in [49], [50]. This causes the real number of COVID-19 cases to match the COVIDHunter's predicted number of cases using a mitigation measure strength of 0.4. This informs us that the mitigation measures are further strengthened from 0.3 to 0.4, which is in line with the actual mitigation measures taken in Switzerland. 2) Even with the increase in the strength of the mitigation measures during November and December 2021, the number of COVID-19 cases keeps increasing after January 2022. We believe this is mainly because of the new variant, Omicron, that starts circulating in the population of Switzerland around the start of December 2021 [51]. 3) The IHME's predicted number of cases also matches that of the FOPH's number of cases. However, this indicates that the IHME model provides an inaccurate prediction (i.e., underestimation) as IHME provides the predicted number of COVID-19 cases assuming the strength of the mitigation measures during November and December 2021 to remain the same as that applied before November 2021, which is incorrect based on governmental information [49], [50]. 4) The ICL model provides a significantly underestimated number of cases even when the ICL model is configured for increased strength of the mitigation measures by 50% (ICL+50%).

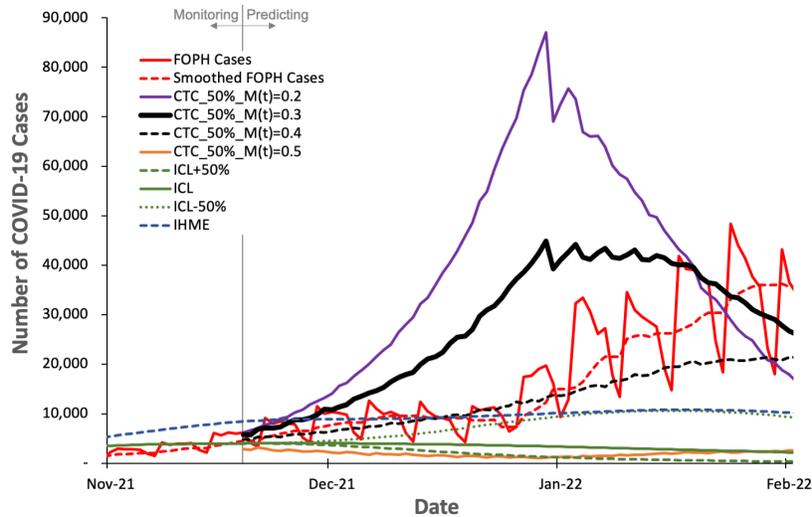

**Figure 7. Predicted number of COVID-19 cases by COVIDHunter model and other two state-of-the-art models, ICL and IHME, compared to the real number (FOPH Cases) of COVID-19 cases released after performing the prediction.**

**Figure 8** shows the number of COVID-19 hospitalizations and deaths predicted by the three models and the real numbers released by FOPH, called "FOPH Hospitalizations" and "FOPH Deaths", respectively. We make four key observations. 1) The FOPH's numbers of COVID-19 hospitalizations and deaths show a certainty rate level of 50%. 2) The COVIDHunter model with a certainty rate level of 50% and an increase in the mitigation measure strength from 0.3 to 0.4 provides an accurate prediction of both the number of hospitalizations and the number of deaths, which is in line with the real numbers provided by FOPH until a new variant, Omicron, is introduced. 3) The Omicron variant, subsequent increases in mitigation measure strength, and increase in vaccination rate cause fewer hospitalizations and deaths than that predicted by COVIDHunter after January 2022. We did not configure COVIDHunter to account for the Omicron variant when we perform the prediction since the Omicron variant was not a variant of concern in November 2021. 4) Similar to the third and fourth observations we make for **Figure 7**, we observe that both ICL and IHME provide inaccurate predictions. That is the ICL model still provides significantly underestimated statistics and IHME provides predictions that match the FOPH's numbers, which we believe is implausible as the circumstances of virus variant, mitigation measures, and vaccination rates during January 2022 are very different from that in November 2021.

We conclude that choosing the appropriate configurations for COVIDHunter leads to accurate predictions of numbers of cases, hospitalizations, and deaths. We demonstrate that COVIDHunter is more accurate than state-of-the-art prediction models, ICL and IHME.

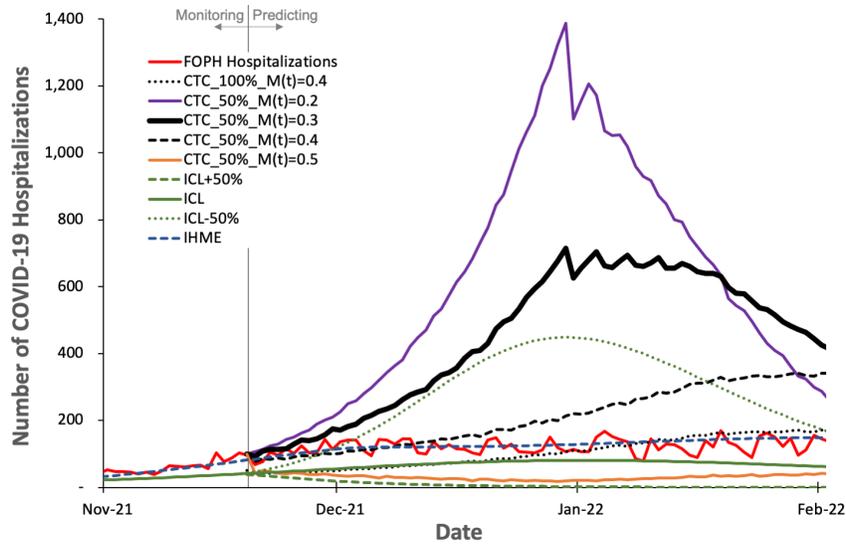

(a)

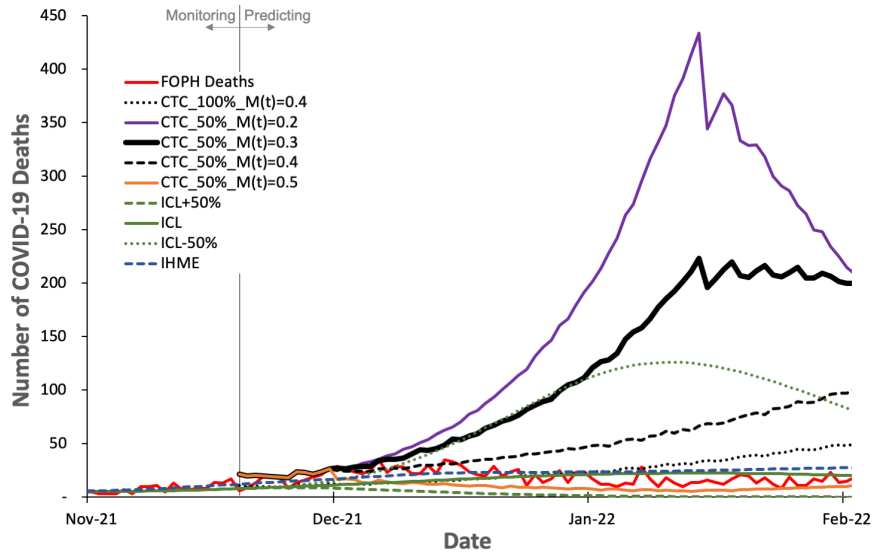

(b)

**Figure 8. Predicted number of COVID-19 (a) hospitalizations and (b) deaths by COVIDHunter, ICL, and IHME, compared to the real numbers (FOPH Hospitalizations and FOPH Deaths) released after performing the prediction.**

## 4. CONCLUSION AND DISCUSSION

We conclude that COVIDHunter provides a more accurate estimation of the number of COVID-19 cases, compared to IHME (which provides inaccurate estimation during the first wave) and ICL (which provides over-estimation), with complete control over the certainty rate level, mitigation measures, and environmental conditions. Unlike LSHTM, COVIDHunter also ensures no prediction delay. We demonstrate the effectiveness of COVIDHunter through about 2 years of monitoring COVID-19 and two prediction runs as we provide in **Section 3** and the **Supplementary Materials (Supplemental Figures S1-S5)**. COVIDHunter gains these unique advantages over existing models by considering environmental conditions, transmissibility of different variants, and vaccination statistics in our modeling.

Using COVIDHunter, we demonstrate that curbing the spread of COVID-19 in Switzerland requires applying stricter mitigation measures than that of the currently applied mitigation measures for at least 30 days. If the authorities maintain the current mitigation measures, we will face another wave that is very similar to the first wave we had in February 2020. Relaxing the mitigation measures should not be an option before at least February 2022. We provide insights on the effect of each change in the strength of the applied mitigation measure on the number of daily cases, hospitalizations, and deaths. We make all the data, statistical analyses, and a well-documented model implementation publicly and freely available to enable full reproducibility and help society and decision-makers.

We especially build COVIDHunter model to be flexible to configure and easy to extend for representing any existing or future scenario using different values of the three terms of **Equation 1**, 1) $R0$, 2) $M(t)$, 3) $C_e(t)$, in addition to several other parameters such as different variants of concern, vaccination rate, population, number of travelers, percentage of expected infected travelers to the total number of travelers, and hospitalizations- or deaths-to-cases ratios. The COVIDHunter model considers each location independently of other locations, but it also accounts for potential movement between locations by adjusting the corresponding parameters for travelers. By allowing most of the parameters to vary in time, $t$, the COVIDHunter model is capable of accounting for any change in transmission intensity due to changes in environmental conditions and mitigation measures over time. The flexibility of configuring the environmental coefficient and mitigation coefficient allows our proposed model to control for location-specific differences in population density, cultural practices, age distribution, and time-variant mitigation responses in each location.

COVIDHunter has three main limitations that can be addressed in future work. 1) Our modeling approach acts across the overall population without assuming any specific age structure for transmission dynamics. It is still *possible* to consider each age group separately using individual runs of COVIDHunter model simulation, each of which has its own parameter values adjusted for the target age group. 2) The current implementation of COVIDHunter considers only two variants of concerns at the time. 3) COVIDHunter does not consider different types of vaccines nor different immunity/protection periods after vaccination. Instead COVIDHunter treats all types of vaccines equally and it considers only the vaccination rate per day.

**Author contributions**

M.A., S.T., and O.M. led the project. N.A. performed the statistical analysis. M.A. and J.K. produced the tables and figures. M.A., J.K., N.A., and S.T. developed the algorithms and created scripts for running and evaluating simulation runs. All authors wrote, reviewed, and edited the manuscript. All authors read and approved the final manuscript.

**Competing interests**

The authors declare no competing interests.

**Funding**

The authors received no financial support for this project.

**Data sharing statement**

All data and code required to reproduce the experimental results covered in the manuscript are freely available on GitHub: https://github.com/CMU-SAFARI/COVIDHunter. We also show how to flexibly configure our model for any scenario and easily extend it for different measures and conditions than we account for.

**Supplementary Materials**

**S1 Evaluated datasets**

Our experimental evaluation uses a large number of different real datasets, including 1) daily *R* number values, 2) observed daily number of COVID-19 cases, 3) observed daily number of COVID-19 hospitalizations, 4) observed daily number of COVID-19 deaths, 5) number of excess deaths, 6) the estimated strength of mitigation measures as calculated by the Oxford Stringency Index, 7) estimation of COVID-19 statistics as calculated by existing state-of-the-art simulation models, ICL, IHME, LSHTM, and IBZ, from seven different sources as we list below. The raw datasets are provided in the GitHub page of this paper (https://github.com/CMU-SAFARI/COVIDHunter/tree/main/Reproduce-Switzerland-Case-Study-Results) and it can be also obtained from the original sources as we list below:

- Observed COVID-19 statistics (R number values and number of cases, hospitalizations, and deaths)
    - Official reports: https://www.covid19.admin.ch/en/overview
    - Smoothed data: https://ourworldindata.org/coronavirus/country/switzerland?country=~CHE
- Excess deaths:
    - Information: https://www.bfs.admin.ch/bfs/en/home/statistics/health/state-health/mortality-causes-death.html
    - Direct link: https://www.bfs.admin.ch/bfs/en/home/statistics/health/state-health/mortality-causes-death.assetdetail.12607335.html
- Oxford Stringency Index
    - https://www.bsg.ox.ac.uk/research/research-projects/coronavirus-government-response-tracker#data
- The Harvard CRW:
    - Information: https://projects.iq.harvard.edu/covid19/home
    - Direct link: https://projects.iq.harvard.edu/covid19/global
- Imperial College London (ICL) Model:
    - Information: https://mrc-ide.github.io/global-lmic-reports/
    - Direct link: https://github.com/mrc-ide/global-lmic-reports/raw/master/data/2021-04-06v7.csv.zip
- Institute for Health Metrics and Evaluation (IHME) Model:
    - Information: https://mrc-ide.github.io/global-lmic-reports/
    - Direct link: http://www.healthdata.org/covid/data-downloads
- The London School of Hygiene & Tropical Medicine (LSHTM) Model:
    - Information: https://cmmid.github.io/topics/covid19/global_cfr_estimates.html
    - Direct link: https://raw.githubusercontent.com/cmmid/cmmid.github.io/master/topics/covid19/reports/under_reporting_estimates/under_ascertainment_estimates.csv
- The Theoretical Biology Group at ETH Zurich (IBZ) Model:
    - Information: https://ibz-shiny.ethz.ch/covid-19-re-international/

○ Direct link: https://github.com/covid-19-Re/dailyRe-Data

**S2 COVIDHunter Prediction Run on 19 April 2021**

We provide a comprehensive analysis of the COVID-19 statistics provided by our COVIDHuntter model, ICL [9], IHME [10], IBZ [11], and LSHTM [7] from the beginning of the COVID-19 outbreak (February 2020) until 19 April 2021. We then provide a prediction run for the period from 19 April 2021 until 1 June 2021, which was carried out on 19 April 2021.

**S2.1 Observed and predicted $R$ number of SARS-CoV-2**

We calculate the predicted $R$ number using our model (**Equation 1**) and compare it to the observed official $R$ number and the $R$ number of two state-of-the-art models, ICL and IBZ, for the two years of 2020 and 2021. We configure COVIDHunter using the following configurations: 1) CTC as environmental condition approach, 2) certainty rate levels of 50% and 100%, and 3) mitigation coefficient values of 0.35 and 0.7. All our scripts are provided on our GitHub page. We consider the mean $R$ number provided by the ICL model. We consider the median $R$ number calculated by the IBZ model based on the observed number of hospitalized patients. IBZ provides the predicted (after 9 April 2021) $R$ number as the mean of the estimates from the last 7 days.

Based on **Figure S1**, we make three key observations. 1) COVIDHunter predicts the changes in $R$ number much (4-13 days) earlier than that predicted by ICL model, which leads to a more accurate prediction. The $R$ number calculated by COVIDHunter (with a certainty rate level of 50%) before 19 April 2021 is on average 1.1× more than that provided by ICL model, IBZ model, and the observed official $R$ number. Using a certainty rate level of 100%, COVIDHunter predicts the $R$ number to be close in value to the observed $R$ number. The $R$ numbers calculated by IBZ model and official authority (observed) are normally not provided for the last two weeks [11][46]. 2) Our model predicts that the current $R$ number is still higher than 1 (1.215 and 1.099 using certainty rate levels of 50% and 100%, respectively) during April 2021. This indicates that the spread of the SARS-CoV-2 virus is still active and it causes an exponential increase in the number of new cases. 3) Our model predicts that if the mitigation measures that are applied nationwide in Switzerland are tightened (*M(t)* increases from 0.55 to 0.7) for only 30 days (19 April to 19 May 2021), then the $R$ number decreases by at least 1.75 × (from 1.215 to 0.691). However, if the mitigation measures are relaxed (*M(t)* drops from 0.55 to 0.35) for only 30 days (19 April to 19 May 2021), then the $R$ number increases by at least 1.23× (from 1.215 to 1.497).

We conclude that COVIDHunter's estimation of the $R$ number is more accurate than that calculated by the ICL and IBZ models, as validated by the currently observed $R$ number.

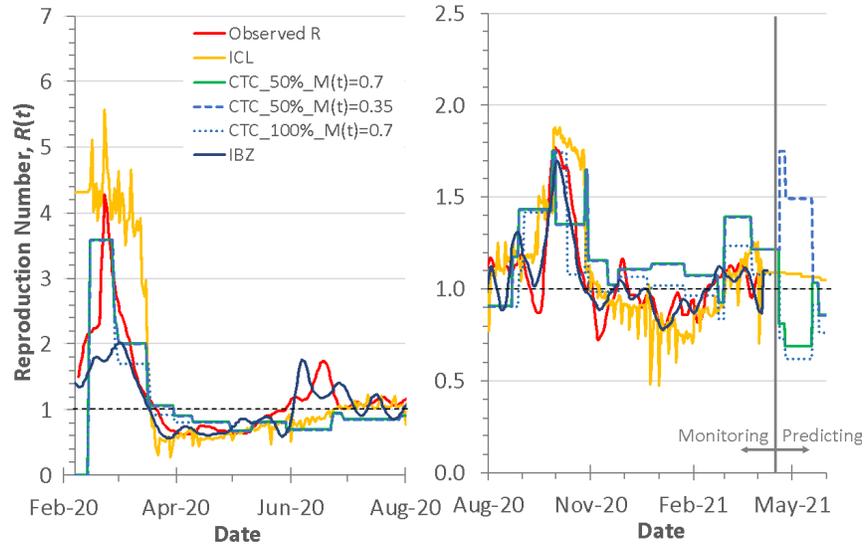

**Figure S1. Observed and predicted reproduction number, $R(t)$, for the two years of 2020 and 2021.** We use CTC environmental condition approach, certainty rate levels of 50% and 100%, and mitigation coefficient values of 0.35 and 0.7 for COVIDHunter. We compare COVIDHunter's predicted $R$ number to the observed $R$ number and two state-of-the-art models, ICL and IBZ. The horizontal dashed line represents $R(t)$ =1.0.

**S2.2 Evaluating the mitigation measures**

We evaluate the mitigation coefficient, $M(t)$, which represents the mitigation measures applied (or to be applied) in Switzerland from January 2020 to June 2021. We use two different environmental condition approaches, CRW and CTC. We assume two certainty rate levels of 50% and 100% to account for uncertainty in the observed number of cases. We use five mitigation coefficients, $M(t)$, values of 0.35, 0.4, 0.5, 0.6, and 0.7 for each configuration of COVIDHunter during 19 April to 19 May 2021. We compare the evaluated mitigation measures to that evaluated by the Oxford Stringency Index, as we provide in **Figure S2**. We also evaluate the mitigation coefficient when we ignore the effect of environmental changes (i.e., by setting $C_e$ =1 in **Equation 1**), while maintaining the same number of COVID-19 cases that provided with a certainty rate level of 50%.

Based on **Figure S2**, we make four key observations. 1) Excluding the effect of environmental changes from the COVIDHunter model, by setting $C_e$ =1 in **Equation 1**, leads to an inaccurate evaluation of the mitigation measures. For example, during the summer of 2020 (between the two major waves of 2020), COVIDHunter (*WithoutCTC_50%*) evaluates the mitigation coefficient to be as high as 0.6. This means that the mitigation measures (*only* mandatory wearing masks on public transport) applied during the summer of 2020 are *only* 14% more relaxed compared to the mitigation measures (e.g., closure of schools, restaurants, and borders, ban on small and large events) applied during the first wave, which is implausible. This highlights the importance of considering the effect of external environmental changes on simulating the spread of COVID-19. Unfortunately, environmental change effects are *not* considered by *any* of the IBZ, LSHTM, ICL, and IHME models, which we believe is a serious shortcoming of these prior models. 2) A drop by 3-30% (as we observe during the mid of November 2020 and the end of August 2020, respectively) in the strength of the mitigation measures for a certain period of time (10 to 20 days) is enough to double the predicted number of COVID-19 cases. 3) We evaluate the strength of the mitigation measures applied in Switzerland to be usually (65% of the time) up to 80% to 131% higher than that provided by the Oxford Stringency Index. 4) The strength of the mitigation measures has changed 11 times and 2 times during the years of 2020 and 2021, respectively, each of which is maintained for at least 9 days and at most 66 days (32 days on average).

We conclude that considering the effect of environmental changes (e.g., daytime temperature) on the spread of COVID-19 improves simulation outcomes and provides an accurate evaluation of the strength of the past and current mitigation measures.

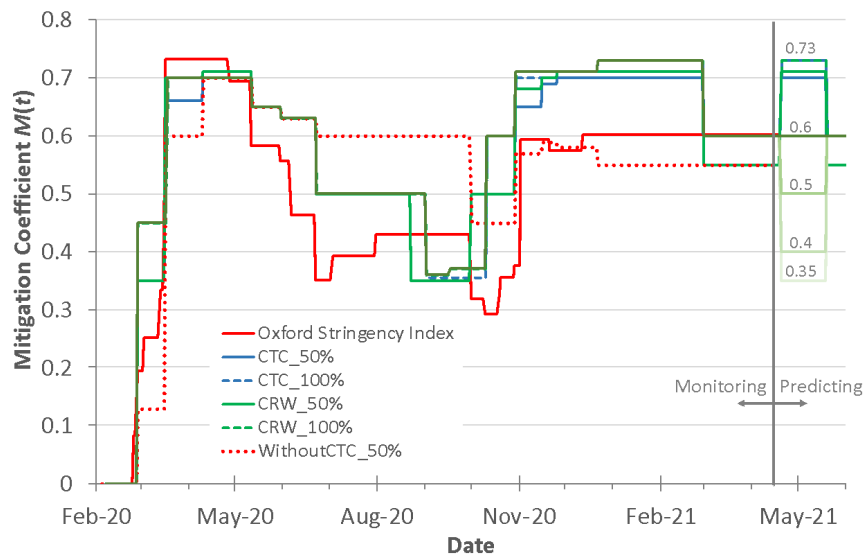

**Figure S2. Predicted strength of the mitigation measures (mitigation coefficient, $M(t)$) applied in Switzerland from January 2020 to May 2021 provided by Oxford Stringency Index and COVIDHunter.** We use two different environmental condition approaches, CRW and CTC. We assume two certainty rate levels of 50% and 100%. We use five mitigation $M(t)$ values of 0.35, 0.4, 0.5, 0.6, and 0.7 for each configuration of our model during 19 April to 19 May 2021. The plot called WithoutCTC_50% represents the evaluation of the current mitigation measures while ignoring the effect of environmental changes.

**S2.3 Evaluating the effect of different mitigation coefficient values on COVIDHunter's predicted number of cases, hospitalizations, and deaths**

Using COVIDHunter, we predict the number of COVID-19 cases, hospitalizations, and deaths from 19 April to 19 May 2021. We show the maximum and the average daily number of COVID-19 cases, hospitalizations, and deaths from 19 April to 19 May 2021 in **Figures S3 and S4**, respectively. We use two environmental condition approaches, CRW and CTC, with a certainty rate level of 50%. We assume five mitigation coefficient, $M(t)$, values of 0.35, 0.4, 0.5, 0.6, and 0.7 for each configuration of COVIDHunter from 19 April to 19 May 2021. This range of mitigation coefficient values covers the lowest (i.e., $M(t)=0.35$) and the highest (i.e., $M(t)=0.7$) strengths of mitigation measures that have been applied during the year 2020.

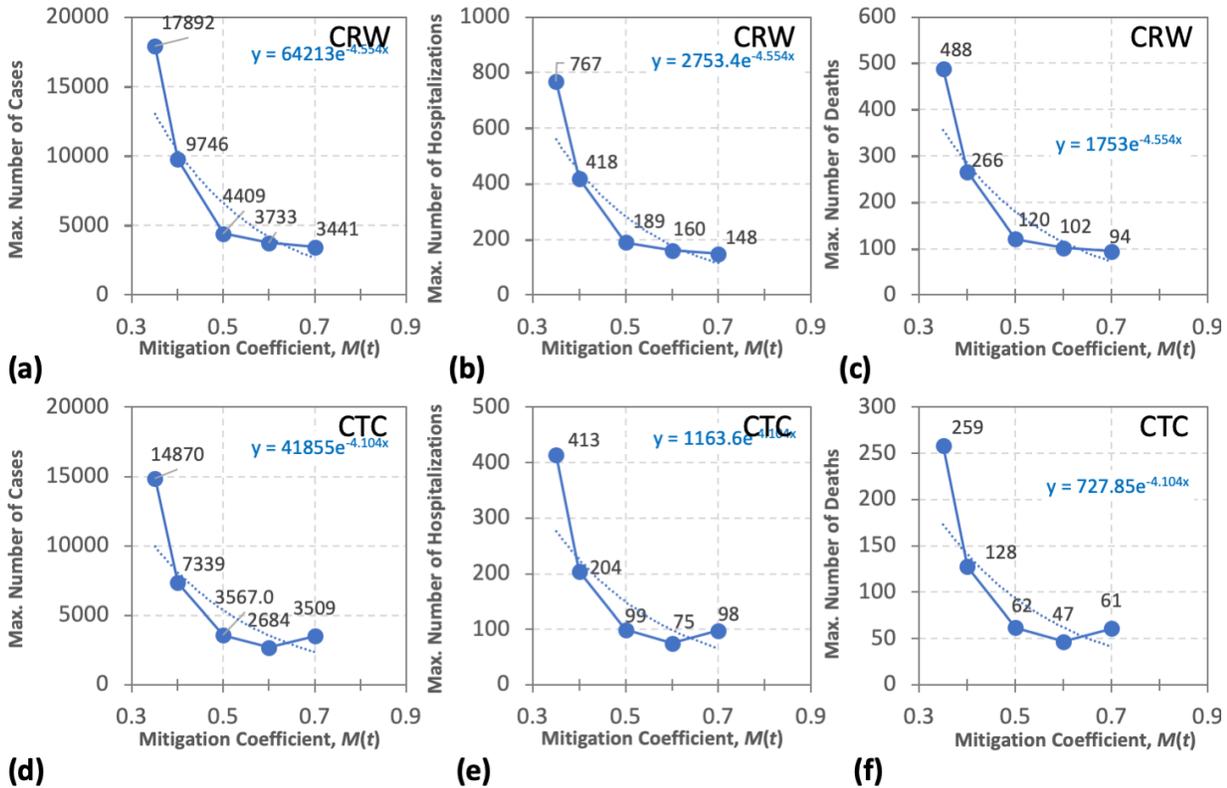

**Figure S3: The maximum daily number of COVID-19 cases, hospitalizations, and deaths as predicted by COVIDHunter from 19 April to 19 May 2021.** We use five mitigation coefficient, $M(t)$, values of 0.35, 0.4, 0.5, 0.6, and 0.7 for each configuration of our model from 19 April to 19 May 2021. We use two different environmental condition approaches, CRW (a)-(c) and CTC (d)-(f) with a certainty rate level of 50%. The dashed line represents an exponential model fit to data.

Based on **Figures S3 and S4**, we make three key observations. 1) COVIDHunter predicts that the maximum daily number of COVID-19 cases, hospitalizations, and deaths from 19 April to 19 May 2021 would be 3441, 148, and 93, respectively, using CRW and $M(t)=0.7$, as we show in **Figure S3(a-c)**. Using our environmental condition approach, CTC, and $M(t)=0.7$, the maximum daily number of COVID-19 cases, hospitalizations, and deaths from 19 April to 19 May 2021 would be 3509, 98, and 61, respectively, as we show in **Figure S3(d-f)**. 2) Relaxing the mitigation measures (M is changed from 0.55 to 0.35) exponentially increases the maximum daily number of cases, hospitalizations, and deaths by 5.1×, reaching up to 17892, 767, and 488, respectively, as predicted by COVIDHunter with the CRW approach (**Figure S3(a-c)**). Using the CTC approach and $M(t)=0.35$, COVIDHunter predicts an exponential increase in the maximum daily number of cases, hospitalizations, and deaths by only 4.13×, as we show in **Figure S3(a-c)**. This is expected as the CTC approach considers only the drop in temperature rather than the average effect of many environmental conditions as the CRW approach does. 3) Relaxing the mitigation measures ($M(t)$ is changed from 0.55 to 0.35) causes the daily number of cases, hospitalizations, and deaths to exponentially increase by an average of 2.8× and 2.5× from 19 April to 19 May 2021 using CRW and CTC environmental approaches, respectively, as we show in **Figure S4**.

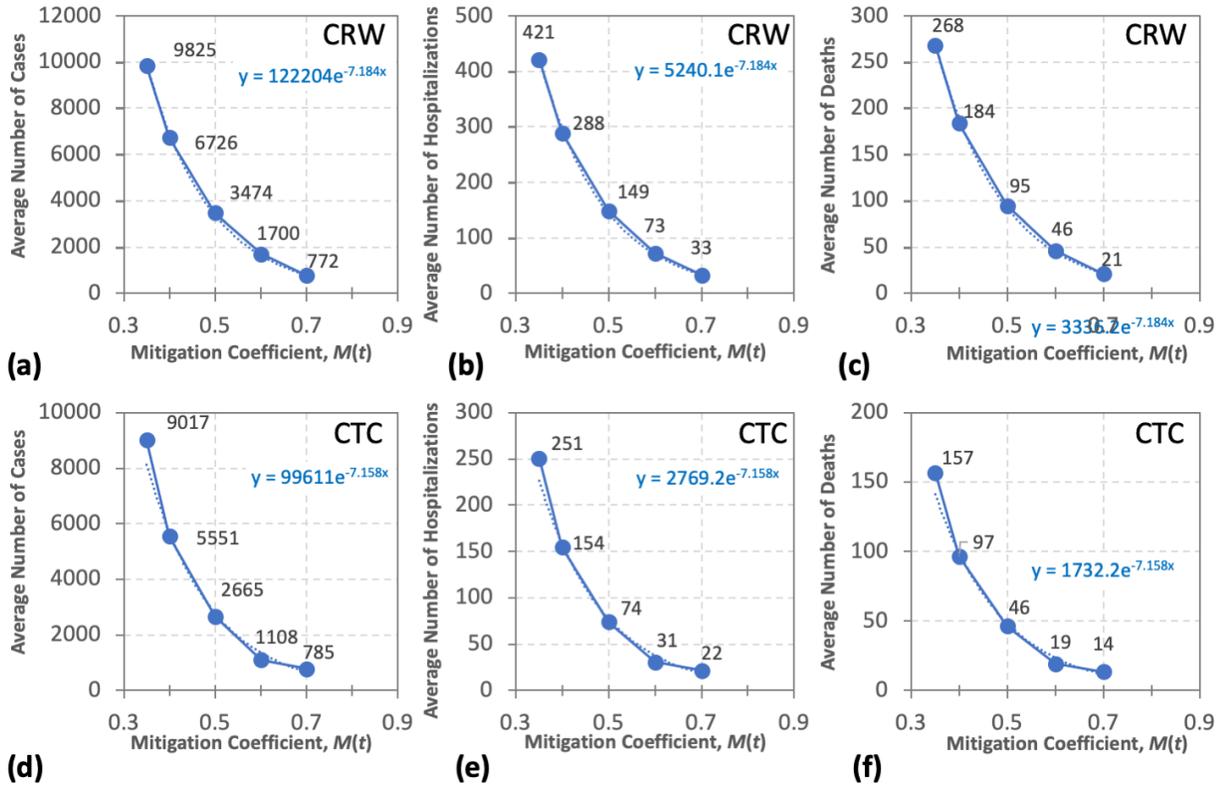

**Figure S4: The average daily number of COVID-19 cases, hospitalizations, and deaths as predicted by COVIDHunter from 19 April to 19 May 2021.** We use five mitigation coefficient, $M(t)$, values of 0.35, 0.4, 0.5, 0.6, and 0.7 for each configuration of our model from 19 April to 19 May 2021. We use two different environmental condition approaches, CRW (a)-(c) and CTC (d)-(f) with a certainty rate level of 50%. The dashed line represents an exponential model fit to data.

We conclude that COVIDHunter provides a flexible evaluation of the effect of different strengths of the past and current mitigation measures on the number of COVID-19 cases, hospitalizations, and deaths. COVIDHunter evaluates the applied mitigation measures with high flexibility in configuring the environmental coefficient and mitigation coefficient, which helps society and decision-makers to accurately review the current situation and estimate future impact of decisions.

**S2.4 Evaluating the predicted number of COVID-19 cases**

We evaluate COVIDHunter's *predicted* daily number of COVID-19 cases in Switzerland. We compare the predicted numbers by our model to the observed numbers and those provided by three state-of-the-art models (ICL, IHME, and LSHTM), as shown in **Figure S5**. We calculate the observed number of cases as the expected number of cases with a certainty rate level of 100%. We use three default configurations for the prediction of the ICL model: 1) strengthening mitigation measures by 50%, 2) maintaining the same mitigation measures, and 3) relaxing mitigation measures by 50% which we refer to as ICL+50%, ICL, and ICL-50%, respectively, in **Figures S5**, **S6**, and **S7**. We use the mean numbers reported by the IHME model that represent the most relaxed mitigation measures, called "no vaccine" by the IHME model. We use the median numbers reported by the LSHTM model.

Based on **Figure S5**, we make four key observations. 1) Our model predicts that the number of COVID-19 cases reduces significantly (less than 50 daily cases) within May 2021 if the mitigation measures that are applied nationwide in Switzerland are tightened (*M(t)* increases from 0.55 to 0.7) for at least 30 days. If the authority decides to relax the mitigation measures to the lowest strength that has been applied during the year 2020 (i.e., $M(t) = 0.35$), then the daily expected number of cases increases by an average of $5.1\times$ and $4.13\times$ (up to 17,892 daily cases) using the CRW and CTC environmental approaches, respectively. 2) COVIDHunter (CTC_100%_M(t)=0.7) predicts the number of COVID-19 cases to be equivalent to that predicted by the IHME model during the second wave with a certainty rate level of 100%. However, during the first wave, the prediction of the IHME model is $3.8\times$ less than the expected number of cases using a certainty rate level of 100%. This means that, unlike our model, the IHME model considers the laboratory-confirmed cases during the first wave to be as if the tests are done at a population-scale, which is very likely incorrect. This is in line with a recent study that demonstrates the high inaccuracy of the IHME model. 3) Overall, our model predicts up to $7.9\times$ and $6.4\times$ (on average $1.9\times$ and $2.1\times$) smaller number of COVID-19 cases than that predicted by ICL model using CTC and CRW approaches, respectively, and a certainty rate of 50%. This suggests that the multiplicative relationship between the confirmed number of cases and the true number of cases can be represented by a certainty rate of 22% to 33%, which our model can easily account for. 4) The number of COVID-19 cases estimated by the LSHTM model during the first wave is 1) on average 24% less than that estimated by COVIDHunter and 2) 10 days late from that predicted by COVIDHunter, IHME, and ICL. The prediction of the LSHTM model during the second wave is not available by the model's pre-computed projections.

We conclude that COVIDHunter provides a more accurate estimation of the number of COVID-19 cases, compared to IHME (which provides inaccurate estimation during the first wave) and ICL (which provides over-estimation), with complete control over the certainty rate level, mitigation measures, and environmental conditions. Unlike LSHTM, COVIDHunter also ensures no prediction delay.

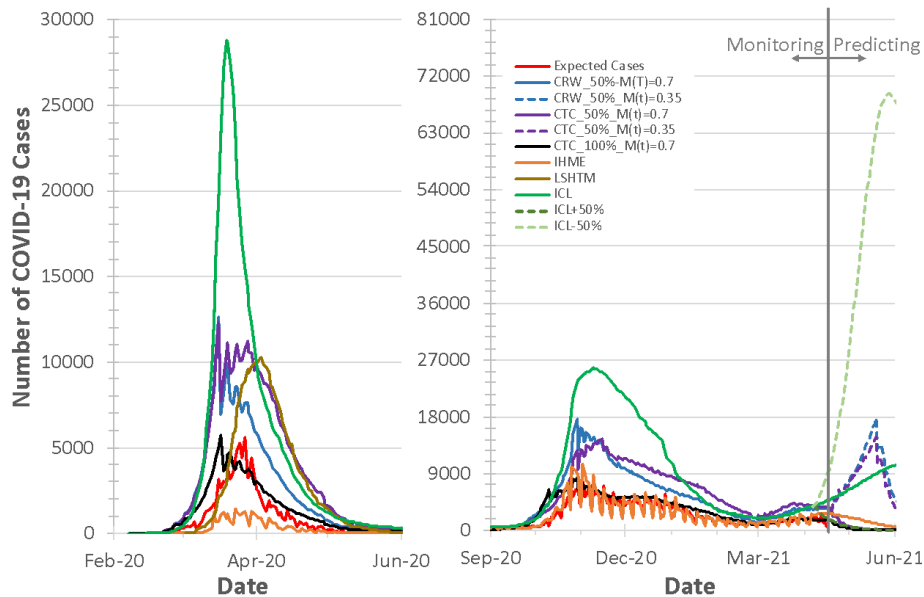

**Figure S5. Observed and predicted number of COVID-19 cases by our model and other three state-of-the-art models.** We use two different environmental condition approaches, CRW and CTC with two certainty rate levels of 50% and 100%. We use two mitigation coefficient, $M(t)$, values of 0.35 and 0.7 for each configuration of our model from 19 April to 19 May 2021.

**S2.5 Evaluating the predicted number of COVID-19 hospitalizations**

We evaluate COVIDHunter's *predicted* daily number of COVID-19 hospitalizations in **Figure S6**. We use the observed official number of hospitalizations as is. Using the number of cases calculated with **Equation 2**, we find $X$ (hospitalizations-to-cases ratio) to be 4.288% and 2.780%, using CRW and CTC, respectively, during the second wave.

We make five key observations based on **Figure S6**. 1) COVIDHunter (CRW_50%_M(t)=0.7) with a certainty rate level of 50% predicts on average $5.33 \times$ smaller number of COVID-19 hospitalizations than that calculated by the IHME model. 2) The ICL model predicts the number of hospitalizations to be similar to that predicted by COVIDHunter (CTC_50%_M(t)=0.7) during the first and the second waves. This suggests that both the ICL model and COVIDHunter (CTC_50%_M(t)=0.7) consider that the actual number of COVID-19 hospitalizations is twice the observed number of COVID-19 hospitalizations. 3) COVIDHunter with a certainty rate level of 100% predicts the number of cases to perfectly fit the curve of the observed number of hospitalizations, reaching up to 231 hospitalized patients a day. 4) Our model predicts that the number of COVID-19 hospitalizations reduces significantly (less than 5 daily hospitalized patients) within May 2021 if the mitigation measures that are applied nationwide in Switzerland are tightened ($M(t)$ increases from 0.55 to 0.7) for at least 30 days. This is in line with what the ICL model (ICL+50%) predicts when ICL model is configured to strengthen the mitigation measures by 50%. If the authority decides to relax the mitigation measures to the lowest strength that has been applied during the year 2020 ($M(t)$ drops from 0.55 to 0.35), then the daily expected number of hospitalizations *exponentially* increases by an average of $5.1\times$ and $4.13\times$, becoming as high as the peak of the second wave (up to 767 daily hospitalized patients), using the CRW and CTC environmental approaches, respectively. ICL model predicts the situation to be worst, showing $2\times$ and $3.74\times$ higher number of hospitalizations than COVIDHunter CRW_50%_M(t)=0.35 and CRW_50%_M(t)=0.35, respectively, when ICL model is configured to 50% relaxation in the mitigation measures. 5) The use of the CTC approach for determining the environmental coefficient value yields a slightly different number (on average $1.7\times$ less) of hospitalizations compared to that provided by the use of the CRW approach. This is expected as the CTC approach considers only the monthly average change in temperature, whereas the CRW approach considers the daily change in *several* environmental conditions.

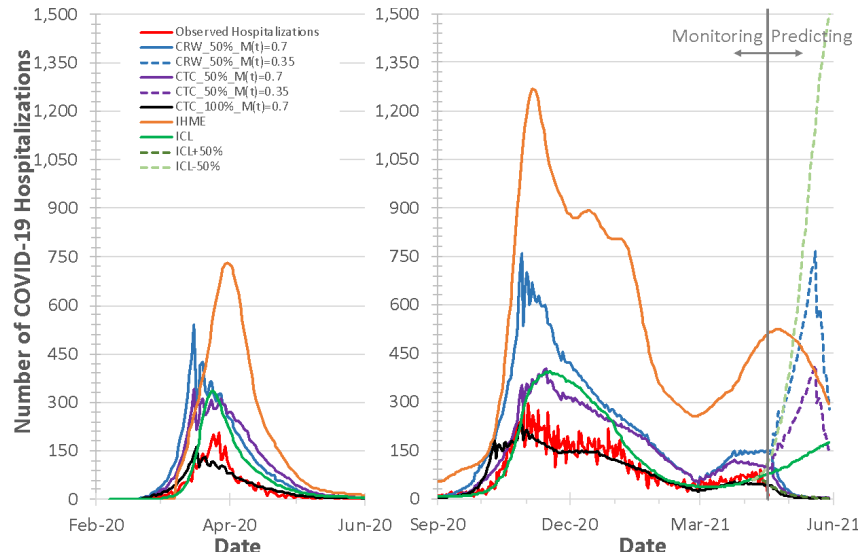

**Figure S6. Observed and predicted number of COVID-19 hospitalizations.** We use two different environmental condition approaches, CRW and CTC with two certainty rate levels of 50% and 100%. We use two mitigation coefficient values, $M(t)$, of 0.35 and 0.7 for each configuration of our model from 19 April to 19 May 2021.

We conclude that 1) unlike the IBZ and LSHTM models, COVIDHunter is able to predict the number of hospitalizations and 2) COVIDHunter provides more accurate estimation of the number of hospitalizations compared to that calculated by ICL (which provides overestimation) and IHME (which provides late estimation). COVIDHunter predicts the number of COVID-19 hospitalizations in a simple, convenient and flexible way that requires calculating only the daily number of cases and the hospitalization-to-cases ratio, $C_X$.

## S2.6 Evaluating the predicted number of COVID-19 deaths

We evaluate COVIDHunter's *predicted* daily number of COVID-19 deaths in **Figure S7** after accounting for the 15-day shift (as we discuss in **Section S3.3**). We calculate the observed number of deaths as the number of excess deaths to account for uncertainty in reporting COVID-19 deaths. Using the number of cases calculated using **Equation 2**, we find $Y$ (deaths-to-cases ratio, using excess death data) to be 2.730% and 1.739%, using CRW and CTC, respectively, during the second wave.

We make three key observations based on **Figure S7**. 1) COVIDHunter with a certainty rate of 100% predicts the number of deaths to perfectly fit the three curves of the observed number of *excess* deaths, ICL deaths, and IHME deaths, reaching up to 144 deaths a day. During the second wave, the ICL curve is shifted (late prediction) by 5-10 days from that of other models. 2) Similar to what we observe for the number of hospitalizations, our model predicts that the number of COVID-19 deaths significantly reduces (reaching up to a single death a day) with stricter mitigation measures ($M(t)$ increases from 0.55 to 0.7) maintained for at least the upcoming 30 days. This is in line with what the IHME model predicts. Relaxing the mitigation measures ($M(t)$ drops from 0.55 to 0.35) *exponentially* increases the death toll by an average of $5.1\times$ and $4.13\times$, reaching up to 488 new daily deaths, as predicted by COVIDHunter using CRW and CTC environmental condition approaches, respectively. COVIDHunter's prediction (CRW_50%_M(t)=0.35) is in line with what ICL model predicts, when ICL model is configured as 50% relaxation in the mitigation measures. 3) During the first wave, the use of a certainty rate of 50% provides $3\times$ and $2.7\times$ ($2.6\times$ and $1.7\times$ during the second wave) higher number of deaths compared to that provided by ICL and IHME models, when COVIDHunter uses CRW and CTC environmental condition approaches, respectively.

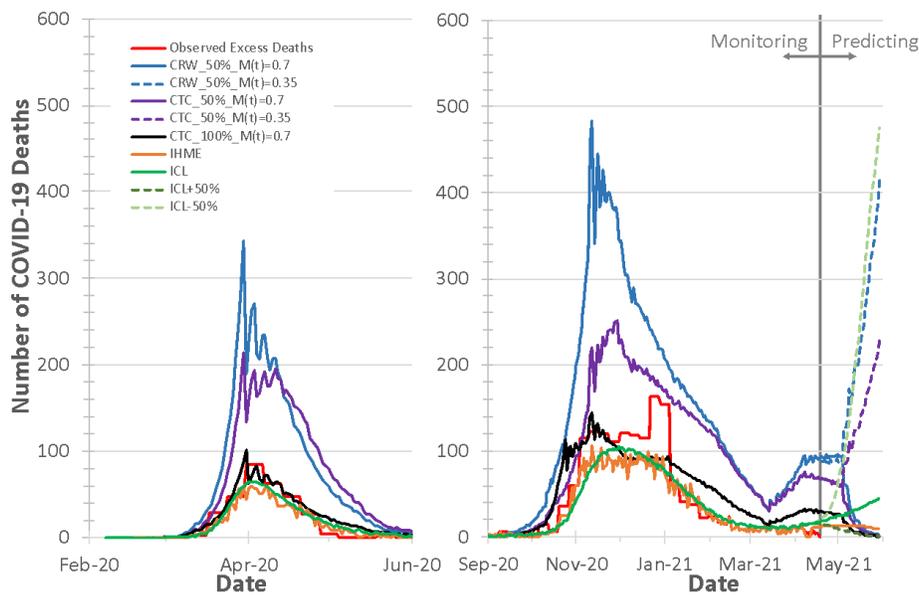

**Figure S7. Observed and predicted number of COVID-19 deaths.** We use two different environmental condition approaches, CRW and CTC with two certainty rate levels of 50% and 100%. We use two mitigation coefficient values, $M(t)$, of 0.35 and 0.7 for each configuration of our model from 19 April to 19 May 2021.

We conclude that 1) unlike the IBZ and LSHTM models, COVIDHunter is able to predict the number of deaths, 2) COVIDHunter predicts the number of deaths to be similar to that predicted by the ICL and IHME models. Yet, COVIDHunter provides a more accurate estimation of other COVID-19 statistics ($R$, number of cases and hospitalizations) compared to ICL and IHME, as we comprehensively evaluate in the previous sections, and 3) COVIDHunter requires calculating only the daily number of cases and the deaths-to-cases ratio, $C_Y$, to predict the daily number of deaths.

**S2.7 Evaluating the effect of different vaccination rates**

We evaluate the effect of different vaccination rates of 0, 0.1, 0.28, 0.4, and 0.5 per day on the reproduction number and the daily number of COVID-19 cases in **Figure S8**. We set the first day of vaccination availability in Switzerland as 28 February 2021 based on governmental data (https://www.covid19.admin.ch/en/overview). We choose May 2021 for our evaluation as it precedes the introduction of the Delta variant in the population of Switzerland and the strength of the mitigation measures remains the same throughout the entire month. This helps us to isolate/reduce potential factors (except the number of vaccinated persons) that can affect the reproduction number and the number of COVID-19 cases. Based on Figure S8, we make three key observations. 1) For each 0.1 rise in the vaccination rate per day, there is on average a 0.07 decrease in the reproduction number, as shown in **Figure S8(a)**. 2) For each 0.1 rise in the vaccination rate per day, there is an exponential decrease in the number of COVID-19 cases, as shown in **Figure S8(b)**.

We conclude that the vaccination rate per day is a key factor that directly affects the average number of new infections caused by each infected person at a given point in time and thus it affects the number of cases.

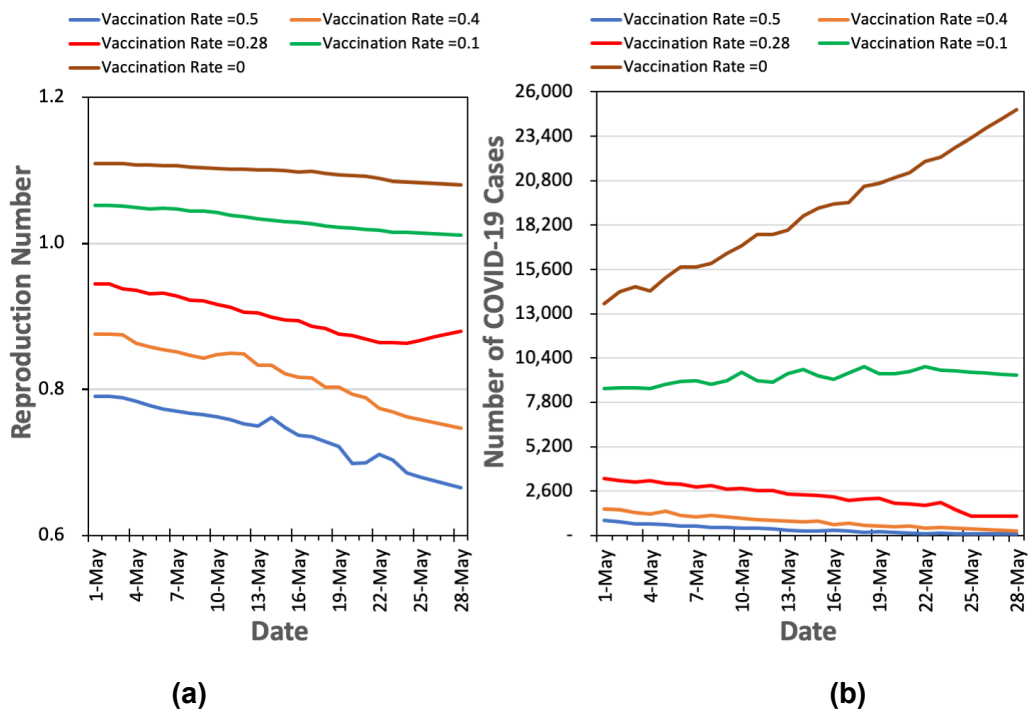

(a)　　　　　　　　　　　　　　　　　　(b)

**Figure S8. (a) The reproduction number and (b) the number of COVID-19 cases calculated by COVIDHunter using different vaccination rates per day during May 2021.** We use the CTC environmental condition approach with a certainty rate level of 50%.